\documentclass[12pt]{article}
\usepackage{xcolor}
\usepackage{graphicx}
\usepackage{amsmath}
\usepackage{amssymb}
\usepackage{geometry}
\usepackage{array, tabularx}
\usepackage{multirow}
\usepackage{pdflscape}
\usepackage{amsfonts}
\usepackage{placeins}
\usepackage{hyperref}
\usepackage{url}
\usepackage{caption}
\usepackage{epstopdf}
\usepackage[normalem]{ulem}
\usepackage{subcaption}
\begin{document}
\title{Risk-Dominant Equilibrium in Quantum Prisoner's Dilemma}

\author{Ahmed S. Elgazzar\\
\textit{Mathematics Department, Faculty of Science, Arish University}\\
\textit{45516 Arish, Egypt}\\
aselgazzar@aru.edu.eg\\
\url{https://orcid.org/0000-0002-6398-4794}}
\date{}
\maketitle

\begin{abstract}
The choice of a unique Nash equilibrium (NE) is crucial in theoretical classical and quantum games. The Eiswer-Wilkens-Lewenstein quantization scheme solves the prisoner's dilemma only for high entanglement. At medium entanglement, there are multiple NEs. We investigate the selection of a unique NE in the quantum prisoner's dilemma with variable dilemma strength parameters. The risk-dominance criterion is used. The influence of the dilemma strength parameters and entanglement is emphasized. We found that entanglement completely controls the risk-dominant equilibrium. Entanglement promotes quantum-cooperation in the risk-dominant equilibrium and thus improves its outcome.
\end{abstract}
\section{Introduction}
Game theory \cite{Neumann44} is a mathematical framework for analyzing the strategic interactions between rational decision-makers. A game consists of a well-defined group of participants and a set of actions available to each participant. Outcomes depend not only on the individual actions, but also on the actions of the others. Game theory is applicable in various disciplines, including economics \cite{appecon}, biology \cite{Colman95,Tanimoto15}, computer science \cite{computer}, political science \cite{political} and psychology \cite{psychology}. In a two-player, two-strategy ($2\times2$) game, there are two participants, each of whom has two possible actions to choose from. This class of games provides a simplified framework for analyzing strategic pairwise interactions and serves as a building block for more complex game structures. An interesting formulation with variable dilemma strength parameters is presented in Refs. \cite{Tanimoto07,Ito18}.

In non-cooperative games \cite{Nash51,Fuj15}, each player makes decisions independently without formal agreements or communication. This situation mirrors many real-life scenarios, such as competitive markets, negotiations and conflicts. The study of non-cooperative games allows us to analyze strategic interactions, understand the incentives and motivations of decision-makers, and predict their behavior.

Nash equilibrium (NE) \cite{Nash50} is a central concept in game theory. It is a state in which the participants have no incentive to deviate from the chosen actions. It predicts how rational, self-interested participants are likely to behave. A Pareto-optimal (PO) equilibrium is a state in which it is impossible to make one participant better off without making another worse off. A PO equilibrium is considered socially desirable because it maximizes group benefit and avoids situations where exploitation or injustice may occur. Therefore, NE focuses on individual incentives, while PO equilibrium emphasizes the overall efficiency of the outcome.

Various dilemmas can arise in non-cooperative games. The prisoner's dilemma (PD) \cite{pd,Ahmedpd05} is a well-known game in which participants face a dilemma between cooperation and defection. The PO outcome results from mutual cooperation, but the unique NE is mutual defection, which leads to a non-optimal outcome for both participants. The PD has often been used to model conflict situations in various fields \cite{pdapp}.

Other dilemmas in non-cooperative games can arise from multiple NEs. The chicken (CH) game \cite{Ch} models a situation in which participants are faced with a conflict between risk-taking and safety, leading to the dilemma of who backs out first to avoid a disastrous outcome.

The stag-hunt (SH) game \cite{Sh} illustrates a conflict between a risky collective benefit and a guaranteed but lower individual benefit. In the SH game, there are two possible symmetric NEs: mutual cooperation and mutual defection. Although mutual cooperation is PO, the dilemma arises from the uncertainty about the other participant's action. If one participant decides to cooperate but the other one defects, the cooperator is left empty-handed.

The battle of the sexes game \cite{Coord} is another $2\times2$ dilemma that presents a scenario in which two individuals have to decide independently on a joint activity. It illustrates the challenge of coordination when the individuals have different preferences. The individuals are faced with the dilemma of choosing between two unfair NEs.

The selection of a unique equilibrium solution is a crucial task. Several methods have been proposed. The focal point method \cite{focal1,focal2} is based on the idea that certain equilibria are particularly important to participants due to their inherent characteristics or cultural context. These characteristics can guide participants to a particular equilibrium without explicit communication. This concept corresponds to decision making in the real world, where social conventions and cultural norms often determine behavior. However, there are some limitations. First, focal points can vary by culture and situation. Second, there may be multiple focal points, or none may be recognizable. Third, focal points do not always lead to the optimal outcome.

The payoff-dominant equilibrium \cite{HS88} offers higher payoffs to all participants compared to any other equilibrium. This is a clear and objective criterion. However, the payoff-dominance criterion focuses exclusively on the payoff amounts and does not take into account the risk preferences of the participants or possible deviations from the equilibrium. The Harsanyi-Selten algorithm \cite{HS88} gives absolute priority to the payoff-dominant criterion. If the payoff-dominant criterion fails, the Harsanyi-Selten algorithm takes the risk-dominance criterion into account.

The risk-dominance criterion \cite{HS88,rde1} compares the risk of each equilibrium, which is usually measured by the potential loss a participant could suffer if the opponent deviates. The equilibrium with the lower risk is called the risk-dominant equilibrium (RDE). Among the various equilibrium selection criteria, risk dominance offers unique advantages. The risk dominance criterion explicitly accounts for uncertainty in non-cooperative games and focuses on equilibria that are less prone to deviation. This corresponds to real-life situations where players are often risk averse and prefer to avoid potential losses. By choosing the RDE, participants can ensure an appropriate payoff even if their opponent deviates from the expected strategy. The RDE tends to be more stable and robust compared to other equilibria. This is because deviations from an RDE are likely to lead to worse outcomes for the deviating participant, which discourages such deviations and strengthens the stability of the RDE. The RDE can provide participants with a reliable focal point. In some cases, the RDE coincides with the PO equilibrium and can improve group benefit \cite{rde1}. The risk-dominance criterion is fairer than other criteria. In addition, many theoretical \cite{Kandori93,Young93,Lee03} and experimental \cite{Experiment00,Experiment02} studies support the selection of the RDE. There are also some limitations. The risk-dominance criterion requires that the players have similar risk aversion. It is difficult to apply the RDE in games with larger strategy spaces. We can overlook these limitations since this study only deals with $2\times2$ games.

Quantum game theory \cite{Meyer,EWL,MW,rev} combines game theory with some principles of quantum information theory \cite{qinfo}, such as superposition and entanglement. The two quantization schemes Eisert-Wilkens-Lewenstein (EWL) \cite{EWL} and Marinatto-Weber (MW) \cite{MW} are among the pioneering works in this field. Each scheme provides a framework for the quantization of a classical game. The players share an entangled quantum state and use quantum strategies that are unitary operators. The superposition principle extends the range of possible actions. Entanglement allows participants to perform actions based on the shared entangled quantum state. Consequently, players can achieve outcomes that are not possible with classical strategies, which can lead to novel NEs. Quantum games can promote cooperation and lead to PO outcomes even in classical games with non-PO NEs. Therefore, many classical dilemmas can be resolved by quantization.

On the other hand, there are also some disadvantages. Quantization can only solve some classical dilemmas in a narrow range of entanglement degrees \cite{Du03}. The quantum solution is sensitive to the corruption in the initial entangled state \cite{corr1,corr2,1p}. Decoherence has been found to reduce the average payoff in quantum games \cite{decoh1,decoh2}. The NE properties of quantum games are completely controlled by the entangled initial state \cite{cooperat,unique,arbiter1,arbiter2,arbitrary}. Moreover, it is unlikely that a classical dilemma can be solved in asymmetric quantum strategy spaces \cite{asymmetric1,asymmetric2}.

There are very few scientific studies dealing with equilibrium selection in quantum games. Fr\c{a}ckiewicz \cite{qbs} applied the Harsanyi-Selten algorithm to the quantum battle of the sexes game with MW scheme. Situ \cite{situ} found that entanglement can improve coordination and outcomes in asymmetric quantum coordination games with MW scheme. Furthermore, Situ \cite{situ} introduced a measure of the risk of a NE that depends on the maximum loss of a player due to the opponent's deviation from this NE.

By focusing on quantum PD (QPD) using the EWL scheme \cite{Du03}, quantization can resolve the dilemma only at high levels of entanglement. At medium levels of entanglement, the QPD can have multiple NEs. At a low degree of entanglement, the QPD reverts to the classical game and the dilemma remains. The aim of this study is to solve the dilemma of choosing a unique NE in the QPD. We apply the Situ approach \cite{situ} and risk-dominance criterion \cite{HS88,rde1} and investigate the effects of the dilemma strength parameters and entanglement.

The rest of this paper is organized as follows. In Section \ref{sec:2}, we study the NE properties of the general symmetric $2\times2$ game and emphasize the effects of the dilemma strength parameters. In Section \ref{sec:3}, we briefly review the Harsanyi-Selten algorithm \cite{HS88} focusing on the risk-dominance criterion. Section \ref{sec:4} is devoted to the QPD using the EWL quantization scheme \cite{EWL,Du03} with the one-parameter strategy space \cite{1p}. In Section \ref{sec:5}, we investigate the dilemma of choosing a unique NE in the QPD. Both the Situ \cite{situ} and the RDE \cite{rde1} approaches are used. The influence of the dilemma strength parameters and entanglement is emphasized. Some conclusions are presented in Section \ref{sec:6}.

\section{Classical $2\times2$ dilemmas} \label{sec:2}
Symmetric $2\times2$ non-cooperative games represent a fundamental aspect of game theory. In these games, there are two players: A and B. Each player has two possible actions from which he/she can choose independently. Let the possible actions be cooperate ($ C $) and defect ($ D $). Let $R$ be the reward for mutual cooperation, and $P$ be the punishment for mutual defection, so that $R>P$. Define the strength of the gamble-intending dilemma \cite{Tanimoto07,Ito18}, $D_{g}$, as the gain due to defection if the opponent cooperates. Conversely, the strength of the risk-averting dilemma \cite{Tanimoto07,Ito18}, $D_{r}$, is the loss due to cooperation if the opponent defects. In order to simplify the calculations, let $R=1$, $P=0$ and $-1\leq D_{g}, D_{r}\leq1$. The payoff matrix is shown in Table \ref{T1}.
\begin{table}[ht]
	\centering
	\caption{Payoff matrix of a $2\times2$ symmetric game with two dilemma strength parameters, $D_{g}$ and $D_{r}$.}\label{T1}
	\begin{tabular}{c|c c}
		\hline
		& B: $ C $ & B: $ D $\\ [1ex]
		\hline
		A: $ C $ & $ (1,1) $ & $ (-D_{r},1+D_{g}) $  \\
		A: $ D $ & $ (1+D_{g},-D_{r}) $ & $ (0,0) $ \\
		\hline
	\end{tabular}
\end{table}

Depending on the values of $D_{g}$ and $D_{r}$, the following dilemmas arise. If both $D_{g}$ and $D_{r}$ are positive, the PD \cite{pd,Ahmedpd05} results. In the PD, there is a unique non-PO NE: $(D,D)$, while $(C,C)$ leads to the PO outcome. The PD represents the conflict between individual self-interest and collective benefit. If $D_{g}>0$ and $D_{r}<0$, the CH dilemma \cite{Ch} arises, in which there are two asymmetric NEs: $(D,C)$ and $(C,D)$. The CH game represents the conflict between courage and self-preservation. Players must weigh the potential benefits of winning against the potential costs of a disastrous outcome. If $D_{g}<0$ and $D_{r}>0$, the SH \cite{Sh} dilemma arises, in which there are two symmetric NEs: a risky PO NE: $(C,C)$ and a safe non-PO NE: $(D,D)$. The players must choose between these NEs. The case where both $D_{g}$ and $D_{r}$ are negative, is trivial (no dilemma). Table \ref{T2} shows the three dilemmas, their pure-strategy NEs and the corresponding payoffs.
\begin{table}[ht]
	\centering
	\caption{Different $2\times2$ dilemma classes that arise depending on the dilemma strength parameters, $ D_{g} $ and $ D_{r} $, their pure-strategy NEs and the corresponding payoffs.} \label{T2}
	\begin{tabular}{c|c c c}
		\hline
		Dilemma strength parameters & Dilemma class & NE & Payoffs\\
		\hline
		$ D_{g} > 0, D_{r} > 0 $ & PD & $(D,D)$ & $(0,0)$ \\
		\hline
        $ D_{g} > 0, D_{r} < 0 $ & CH & \begin{tabular}{c}
		$(D,C)$ \\
		$(C,D)$ \\
	\end{tabular}
  & \begin{tabular}{c}
		$(1+D_{g},-D_{r})$ \\
		$(-D_{r},1+D_{g})$ \\
	\end{tabular} \\
		\hline
        $ D_{g} < 0, D_{r} > 0 $ & SH & \begin{tabular}{c}
		$(D,D)$ \\
		$(C,C)$ \\
	\end{tabular}
  & \begin{tabular}{c}
		$(0,0)$ \\
		$(1,1)$ \\
	\end{tabular} \\
	\hline
\end{tabular}
\end{table}

In a mixed-strategy game, players choose their actions randomly according to certain probabilities. Suppose player A(B) chooses $C$ with probability $p(q)$. In this case, the expected payoff functions are calculated according to Eqs. (\ref{1}) and (\ref{2}).
\begin{equation}\label{1}
 \emph{\$}^{c}_{\mathrm{A}}(p,q)=(D_{r}-D_{g})pq - D_{r} p + (1+D_{g})q.
 \end{equation}
 \begin{equation}\label{2}
 \emph{\$}^{c}_{\mathrm{B}}(p,q)=(D_{r}-D_{g})pq - D_{r} q + (1+D_{g})p.
 \end{equation}
In a mixed-strategy NE, each player chooses the probability distribution that maximizes his/her expected payoff. Equilibrium is reached when none of the players has an incentive to deviate from the chosen probability distribution. A mixed-strategy NE, $(p^{*},q^{*})$ satisfies Eqs. (\ref{3}) and (\ref{4}).
\begin{equation}\label{3}
\emph{\$}^{c}_{\mathrm{A}}(p^{*},q^{*})\geqslant \emph{\$}^{c}_{\mathrm{A}}(p,q^{*}) \forall p\in[0,1].
\end{equation}
\begin{equation}\label{4}
\emph{\$}^{c}_{\mathrm{B}}(p^{*},q^{*})\geqslant \emph{\$}^{c}_{\mathrm{B}}(p^{*},q) \forall q\in[0,1].
\end{equation}

For the PD, only the probability distribution $(0,0)$ satisfies Eqs. (\ref{3}) and (\ref{4}). Then the dilemma also exists for the mixed-strategy PD. In addition to its many advantages, quantum games \cite{Meyer,EWL,MW,rev} can provide a reasonable solution, see Section \ref{sec:4}. In both CH and SH games, we need to choose a unique NE. This is the topic of the next section.

\section{Equilibrium selection} \label{sec:3}
As can be clearly seen in Table \ref{T2}, both CH and SH games have multiple equilibria. Choosing a unique equilibrium is crucial in games with multiple equilibria. Equilibrium selection techniques determine which NE is more likely to be played. Harasanyi and Selten \cite{HS88} introduced two criteria: payoff-dominance and risk-dominance. Although the payoff-dominant equilibrium can be more advantageous for both players, it is always risky if one player deviates from it. Due to its many advantages, we focus here on the risk-dominance criterion.

\begin{table}[ht]
	\centering
	\caption{Payoff matrix of a general $2\times2$ game with two NEs: $(C,C)$ and $(D,D)$.}\label{T3}
	\begin{tabular}{c|c c}
		\hline
		& B: $ C $ & B: $ D $\\ [1ex]
		\hline
		A: $ C $ & $ (a_{11},b_{11}) $ & $ (a_{12},b_{12}) $  \\
		A: $ D $ & $ (a_{21},b_{21}) $ & $ (a_{22},b_{22}) $ \\
		\hline
	\end{tabular}
\end{table}

In order to illustrate the risk-dominance criterion, let us consider a general $2\times2$ game as shown in Table \ref{T3}. Let $(C,C)$ and $(D,D)$ be two possible NEs. The loss of player A due to player B's deviation from $(C,C)$ is $\Delta_{\mathrm{A}}(C,C)=a_{11}-a_{21}$. Similarly, $\Delta_{\mathrm{B}}(C,C)=b_{11}-b_{12}$. The total loss due to deviation from $(C,C)$ is $\Delta (C,C)=\Delta_{\mathrm{A}}(C,C) \Delta_{\mathrm{B}}(C,C)$. For $(D,D)$, $\Delta_{\mathrm{A}}(D,D)=a_{22}-a_{12}$, $\Delta_{\mathrm{B}}(D,D)=b_{22}-b_{21}$, and $\Delta (D,D)=\Delta_{\mathrm{A}}(D,D) \Delta_{\mathrm{B}}(D,D)$. Equation (\ref{5}) identifies the RDE of this game.
\begin{equation}\label{5}
\mathrm{RDE}=\left\{\begin{array}{ll}
(C,C), & \mathrm{if } \Delta (C,C)>\Delta (D,D); \\
(p^{*},q^{*}), & \mathrm{if } \Delta (C,C)=\Delta (D,D); \\
(D,D), & \mathrm{if } \Delta (C,C)<\Delta (D,D),
\end{array}
\right.
\end{equation}
where
\begin{equation}\label{6}
p^{*}=\frac{\Delta_{\mathrm{B}}(D,D)}{\Delta_{\mathrm{B}}(C,C)+\Delta_{\mathrm{B}}(D,D)}, q^{*}=\frac{\Delta_{\mathrm{A}}(D,D)}{\Delta_{\mathrm{A}}(C,C)+\Delta_{\mathrm{A}}(D,D)}.
\end{equation}

If $(C,D)$ and $(D,C)$ are the two possible NEs, then $\Delta_{\mathrm{A}}(C,D)=a_{12}-a_{22}$, $\Delta_{\mathrm{B}}(C,D)=b_{12}-b_{11}$, $\Delta_{\mathrm{A}}(D,C)=a_{21}-a_{11}$ and $\Delta_{\mathrm{B}}(D,C)=b_{21}-b_{22}$. Equation (\ref{7}) identifies the RDE in this case.
\begin{equation}\label{7}
\mathrm{RDE}=\left\{\begin{array}{ll}
(C,D), & \mathrm{if } \Delta (C,D)>\Delta (D,C); \\
(\acute{p}^{*},\acute{q}^{*}), & \mathrm{if } \Delta (C,D)=\Delta (D,C); \\
(D,C), & \mathrm{if } \Delta (C,D)<\Delta (D,C),
\end{array}
\right.
\end{equation}
where
\begin{equation}\label{8}
\acute{p}^{*}=\frac{\Delta_{\mathrm{B}}(D,C)}{\Delta_{\mathrm{B}}(C,D)+\Delta_{\mathrm{B}}(D,C)}, \acute{q}^{*}=\frac{\Delta_{\mathrm{A}}(C,D)}{\Delta_{\mathrm{A}}(D,C)+\Delta_{\mathrm{A}}(C,D)}.
\end{equation}

To select a unique equilibrium in both the CH and SH games, we apply the risk-dominance criterion \cite{HS88,rde1}. For the CH game, we have
\begin{equation}\label{81}
\Delta(D,C)=\Delta(C,D)=-D_{r}D_{g}.
\end{equation}
Based on the risk-dominance criterion (Eq. (\ref{7})), the RDE is the mixed-strategy equilibrium: $(\frac{-D_{r}}{-D_{r}+D_{g}},\frac{-D_{r}}{-D_{r}+D_{g}})$. For the SH game, $\Delta(C,C)=(-D_{g})^2$ and $\Delta(D,D)=D^{2}_{r}$. According to Eq. (\ref{5}),
\begin{equation}\label{82}
\mathrm{RDE}=\left\{\begin{array}{ll}
(C,C), & \mathrm{if } |D_{g}|>D_{r}; \\
(0.5,0.5), & \mathrm{if } |D_{g}|=D_{r}; \\
(D,D), & \mathrm{if } |D_{g}|<D_{r}.
\end{array}
\right.
\end{equation}
These results were applied to some theoretical game models of Iran-Iraq conflicts over shared resources \cite{rde1}, and reasonable solutions were found.

\section{Quantum PD} \label{sec:4}
In the EWL quantization scheme \cite{EWL,Du03}, we assign a quantum state either: $|C\rangle=\left(
\begin{array}{c}
1 \\
0 \\
\end{array}
\right)
$ or $|D\rangle=\left(
\begin{array}{c}
0 \\
1 \\
\end{array}
\right)
$ to each player. Assume that the game begins with the following initial entangled quantum state
\begin{equation}\label{9}
|\psi_{i}\rangle=\cos(\frac{\gamma}{2})|CC\rangle+\imath\sin(\frac{\gamma}{2})|DD\rangle,
\end{equation}
where $0\leq\gamma\leq\frac{\pi}{2}$ is the measure of entanglement. If $ \gamma = 0 $, $|\psi_{i}\rangle$ is separable, and the game reverts to the classical mixed-strategy PD. If $ \gamma = \frac{\pi}{2} $, the game becomes maximally entangled \cite{EWL}.

Each player is assigned a quantum strategy space, $\hat{U}$. In this study, we use a one-parameter strategic space \cite{1p}, Eqs. (\ref{10}) and (\ref{11}).
\begin{equation}\label{10}
\hat{U}_{\mathrm{A}}(p)=\left\{\begin{pmatrix}
\imath \sqrt{p} & \sqrt{1-p} \\
-\sqrt{1-p} & -\imath \sqrt{p} \\
\end{pmatrix}, 0\leq p \leq1\right\}.
\end{equation}
\begin{equation}\label{11}
\hat{U}_{\mathrm{B}}(q)=\left\{\begin{pmatrix}
\imath \sqrt{q} & \sqrt{1-q} \\
-\sqrt{1-q} & -\imath \sqrt{q} \\
\end{pmatrix}, 0\leq q \leq1\right\}.
\end{equation}
This strategic space is a superposition of the quantum strategies, quantum-cooperate $(\hat{Q}=\hat{U}(1))$ and defect $(\hat{D}=\hat{U}(0))$. This means that player A(B) quantum-cooperates with a probability $p(q)$ and defects with a probability $1-p(1-q)$. The players apply their quantum actions, then a disentangling gate is applied, and the final state of the game, $|\psi_{f}\rangle$ is obtained. For a detailed procedure, see \cite{1p}.

The joint quantum probability distribution is calculated as follows
\begin{equation}\label{12}
\begin{split}
& \epsilon_{1}={\lvert\langle CC|\psi_{f}\rangle\rvert}^{2}=pq, \epsilon_{2}={\lvert\langle CD|\psi_{f}\rangle\rvert}^{2}=p(1-q)\cos^{2}\gamma+(1-p)q\sin^{2}\gamma, \\ & \epsilon_{3}={\lvert\langle DC|\psi_{f}\rangle\rvert}^{2}=(1-p)q\cos^{2}\gamma+p(1-q)\sin^{2}\gamma, \epsilon_{4}={\lvert\langle DD|\psi_{f}\rangle\rvert}^{2}=(1-p)(1-q).
\end{split}
\end{equation}
Using the payoff matrix (Table \ref{T1}) and Eq. (\ref{12}), the expected payoff functions are calculated and expressed in Eqs. (\ref{13}) and (\ref{14}).
\begin{equation}\label{13}
\emph{\$}^{q}_{\mathrm{A}}(\hat{U}_{\mathrm{A}}\otimes\hat{U}_{\mathrm{B}})= (D_{r}-D_{g})pq-D_{r}p+(1+D_{g})q+(1+D_{r}+D_{g})(p-q)\sin^2\gamma.
\end{equation}
\begin{equation}\label{14}
\emph{\$}^{q}_{\mathrm{B}}(\hat{U}_{\mathrm{A}}\otimes\hat{U}_{\mathrm{B}})= (D_{r}-D_{g})pq-D_{r}q+(1+D_{g})p-(1+D_{r}+D_{g})(p-q)\sin^2\gamma.
\end{equation}
The payoffs of the pure quantum strategies can be written in a matrix form, see Table \ref{T4}. The properties of the pure-quantum-strategy NE are summarized in Table \ref{T5}.
\begin{table}[ht]
	\centering
	\caption{Pure-quantum-strategy payoff matrix of the QPD using the one-parameter strategic space, Eqs. (\ref{10}) and (\ref{11}), where $\pi_{\hat{Q}}=-D_{r}+(1+D_{r}+D_{g})\sin^2\gamma$ and $\pi_{\hat{D}}=1+D_{g}-(1+D_{r}+D_{g})\sin^2\gamma$.}\label{T4}
	\begin{tabular}{c|c c}
		\hline
		& B: $\hat{Q}$ & B: $\hat{D}$\\ [1ex]
		\hline
		A: $\hat{Q}$ & $ (1,1) $ & $ (\pi_{\hat{Q}},\pi_{\hat{D}}) $  \\
		A: $\hat{D}$ & $ (\pi_{\hat{D}},\pi_{\hat{Q}}) $ & $ (0,0) $ \\
		\hline
	\end{tabular}
\end{table}

\begin{table}[ht]
	\begin{center}
		\caption{Possible pure-quantum-strategy NEs of the QPD and the corresponding payoffs, where $ \gamma_{1}= \arcsin(\sqrt{\frac{D_{r}}{1+D_{r}+D_{g}}}) $ and $ \gamma_{2}= \arcsin(\sqrt{\frac{D_{g}}{1+D_{r}+D_{g}}}) $.}\label{T5}
		\begin{tabular}{c|c|c|c}
			\hline
			Dilemma strength parameters & Entanglement & NE & Payoff \\ \hline
			$ D_{g}>D_{r} $ &
			\begin{tabular}{c}
				 $0\leq\gamma\leq\gamma_1$ \\
				 \\
				 $\gamma_1\leq\gamma\leq\gamma_2$\\
				 \\
				 $\gamma_2\leq\gamma\leq\frac{\pi}{2}$ \\
			\end{tabular} &
			\begin{tabular}{c}
				$ \hat{D}\otimes \hat{D}$ \\
                $ \hat{D}\otimes \hat{Q}$ \\
				$ \hat{Q}\otimes \hat{D}$ \\
				$ \hat{Q}\otimes \hat{Q}$ \\
			\end{tabular} &
			\begin{tabular}{c}
				$(0,0)$ \\
				$(\pi_{\hat{D}},\pi_{\hat{Q}})$\\
				$(\pi_{\hat{Q}},\pi_{\hat{D}})$\\
				$(1,1)$ \\
			\end{tabular}\\ \hline
			$ D_{g}=D_{r} $ &
			\begin{tabular}{c}
				$0\leq\gamma\leq\gamma_1$ \\
				$\gamma_1\leq\gamma\leq\frac{\pi}{2}$ \\
			\end{tabular} &
			\begin{tabular}{c}
				$ \hat{D}\otimes \hat{D}$ \\
				$ \hat{Q}\otimes \hat{Q}$ \\
			\end{tabular} &
			\begin{tabular}{c}
				$(0,0)$ \\
				$(1,1)$ \\
			\end{tabular}\\ \hline
			$ D_{g}<D_{r} $ &
			\begin{tabular}{c}
				$0\leq\gamma\leq\gamma_2$  \\
				\\
				$\gamma_2\leq\gamma\leq\gamma_1$ \\
				\\
				$\gamma_1\leq\gamma\leq\frac{\pi}{2}$  \\
			\end{tabular} &
			\begin{tabular}{c}
				$ \hat{D}\otimes \hat{D}$ \\
				\begin{tabular}{c}
					$ \hat{D}\otimes \hat{D}$ \\
					$ \hat{Q}\otimes \hat{Q}$ \\
				\end{tabular}\\
				$ \hat{Q}\otimes \hat{Q}$ \\
			\end{tabular} &
			\begin{tabular}{c}
				$(0,0)$ \\
				\begin{tabular}{c}
				$(0,0)$ \\
				$(1,1)$ \\
			    \end{tabular}\\
				$(1,1)$ \\
			\end{tabular}\\ \hline
		\end{tabular}
	\end{center}
\end{table}

As shown in Table \ref{T5}, the dilemma is resolved only when $\gamma_1\leq\gamma\leq\frac{\pi}{2}$ if $ D_{g}\leq D_{r} $, and when $\gamma_2\leq\gamma\leq\frac{\pi}{2}$ if $ D_{g}> D_{r} $. When $ D_{g}>D_{r} $ and $\gamma_1\leq\gamma\leq\gamma_2$ (transitional phase), players face the dilemma of having to choose between two unfair NEs: $ \hat{D}\otimes \hat{Q}$ and $ \hat{Q}\otimes \hat{D}$. This situation is similar to the CH game. When $ D_{g}<D_{r} $ and $\gamma_2\leq\gamma\leq\gamma_1$ (coexistence phase), the players face a dilemma in choosing between two symmetric NEs: $ \hat{D}\otimes \hat{D}$ and $ \hat{Q}\otimes \hat{Q}$. This dilemma is similar to the SH game. The next section deals with these dilemmas using the RDE.

\section{RDE in QPD} \label{sec:5}
In this section, we investigate the dilemma of choosing a unique NE in both the transitional and coexistence phases using both the Situ approach \cite{situ} and the risk-dominance criterion \cite{HS88,rde1}. We also examine the influence of the dilemma strength parameters and entanglement.

\subsection{The transitional phase}
If $ D_{g}>D_{r} $, there are two asymmetric NEs: $ \hat{D}\otimes \hat{Q}$ and $ \hat{Q}\otimes \hat{D}$ in the interval $[\gamma_1,\gamma_2]$. It is clear that $\pi_{\hat{D}}>\pi_{\hat{Q}}$, then choosing one of these NEs is unfair. This situation would tempt both players to defect obtaining the worst payoff $(0,0)$. In this dilemma, rational players would prefer to choose a safer equilibrium.

Situ \cite{situ} has introduced a measure of the risk of a NE, $\delta$. It is the maximum loss of the player who chooses a particular NE due to the opponent's deviation from that particular NE. The risks of the NE: $ \hat{D}\otimes \hat{Q}$ for A and B are given in Eqs. (\ref{15}) and (\ref{16}), respectively.
\begin{equation}\label{15}
\delta_{\mathrm{A}}(\hat{D}\otimes \hat{Q})=\max_{q}[\emph{\$}^{q}_{\mathrm{A}}(\hat{D}\otimes \hat{Q})-\emph{\$}^{q}_{\mathrm{A}}(\hat{D}\otimes \hat{U}_{\mathrm{B}}(q)]= 1+D_{g}-(1+D_{r}+D_{g})\sin^2\gamma.
\end{equation}
\begin{equation}\label{16}
\delta_{\mathrm{B}}(\hat{D}\otimes \hat{Q})=\max_{p}[\emph{\$}^{q}_{\mathrm{B}}(\hat{D}\otimes \hat{Q})-\emph{\$}^{q}_{\mathrm{B}}(\hat{U}_{\mathrm{A}}(p)\otimes \hat{Q})]=0.
\end{equation}
Due to the symmetry of the game, then $\delta_{\mathrm{A}}(\hat{Q}\otimes \hat{D})=\delta_{\mathrm{B}}(\hat{D}\otimes \hat{Q})$ and $\delta_{\mathrm{B}}(\hat{Q}\otimes \hat{D})=\delta_{\mathrm{A}}(\hat{D}\otimes \hat{Q})$. Thus, with the Situ approach, there is no single NE that is safe for both players.

Now we apply the risk-dominance criterion \cite{HS88,rde1}. The total losses due to deviation from both $\hat{D}\otimes \hat{Q}$ and $\hat{Q}\otimes \hat{D}$ are given in Eq. (\ref{17}).
\begin{equation}\label{17}
\Delta (\hat{D}\otimes \hat{Q})=\Delta (\hat{Q}\otimes \hat{D})=[D_{g}-(1+D_{r}+D_{g})\sin^2\gamma][-D_{r}+(1+D_{r}+D_{g})\sin^2\gamma].
\end{equation}
Based on the risk-dominance criterion (Eq. (\ref{7})), the RDE is $\hat{U}_{\mathrm{A}}(p^{*})\otimes\hat{U}_{\mathrm{B}}(q^{*})$, where
\begin{equation}\label{18}
p^{*}=q^{*}=\frac{-D_{r}+(1+D_{r}+D_{g})\sin^2\gamma}{D_{g}-D_{r}}.
\end{equation}
This equilibrium point corresponds to simultaneous quantum-cooperation with probability $p^{*}$ and defection with probability $1-p^{*}$. The probability of quantum-cooperation, $p^{*}$ increases continuously from $0$ at $\gamma=\gamma_1$ to $1$ at $\gamma=\gamma_2$. This increase is faster if the difference $D_g - D_r$ is smaller, and vice versa.

To determine how changes in the parameters $D_{g}$, $D_{r}$ and $\gamma$ affect the degree of quantum-cooperation, we perform a sensitivity analysis on $p^{*}$. In this analysis, we investigate $\frac{\partial p^{*}}{\partial x}$ and calculate the sensitivity index, $S_{x}=\frac{\partial p^{*}}{\partial x}\frac{x}{p^{*}}$, $x=D_{g}, D_{r}$ and $\gamma$. Firstly, we determine the sensitivity of $p^{*}$ to the variation of $D_{g}$.
\begin{equation}\label{19}
\frac{\partial p^{*}}{\partial D_{g}}=\frac{D_{r}-(1+2D_{r})\sin^2\gamma}{(D_{g}-D_{r})^2}.
\end{equation}
If $\frac{\partial p^{*}}{\partial D_{g}}=0$, the critical point for $D_{g}$ is obtained. Then
\begin{equation}\label{20}
\gamma_{g}=\arcsin(\sqrt{\frac{D_{r}}{1+2D_{r}}}).
\end{equation}
If $\gamma_1\leq\gamma<\gamma_{g}$, $\frac{\partial p^{*}}{\partial D_{g}}>0$, then $p^{*}$ increases with increasing $D_{g}$. Conversely, when $\gamma_{g}<\gamma\leq\gamma_2$, $\frac{\partial p^{*}}{\partial D_{g}}<0$, then $p^{*}$ decreases as $D_{g}$ increases.

Secondly, with respect to $D_{r}$, we have
\begin{equation}\label{21}
\frac{\partial p^{*}}{\partial D_{r}}=\frac{-D_{g}+(1+2D_{g})\sin^2\gamma}{(D_{g}-D_{r})^2}.
\end{equation}
The critical point for $D_{r}$ is
\begin{equation}\label{22}
\gamma_{r}=\arcsin(\sqrt{\frac{D_{g}}{1+2D_{g}}}).
\end{equation}
If $\gamma_1\leq\gamma<\gamma_{r}$, $\frac{\partial p^{*}}{\partial D_{r}}$ is negative, then $p^{*}$ decreases as $D_{r}$ increases. Conversely, if $\gamma_{r}<\gamma\leq\gamma_2$, $\frac{\partial p^{*}}{\partial D_{r}}$ becomes positive, then $p^{*}$ increases as $D_{r}$ increases.

Thirdly, the sensitivity of $p^{*}$ to the variation of $\gamma$ is determined by the calculation of
\begin{equation}\label{23}
\frac{\partial p^{*}}{\partial \gamma}=\frac{2(1+D_{g}+D_{r})\sin\gamma\cos\gamma}{D_{g}-D_{r}}.
\end{equation}
In the transitional phase, $\frac{\partial p^{*}}{\partial \gamma}$ is always positive, then $p^{*}$ always increases with $\gamma$. For given values of $D_{g}$ and $D_{r}$, the degree of quantum-cooperation of the RDE increases with entanglement and reaches full quantum-cooperation at $\gamma_2$.

The results of the sensitivity indices of $p^*$ with respect to the parameters $D_{g}$, $D_{r}$ and $\gamma$ are summarized in Table \ref{T6}. It is clear that $D_g$ has a moderate positive (negative) impact on $p^*$ when $\gamma_1\leq\gamma<\gamma_g$ ($\gamma_{g}<\gamma\leq\gamma_2$). If $\gamma_1\leq\gamma<\gamma_r$ ($\gamma_{r}<\gamma\leq\gamma_2$), $D_r$ has a weak negative (positive) influence on $p^*$. Changes in $D_r$ do not significantly affect $p^*$. The degree of entanglement, $\gamma$ has a strong positive influence on $p^*$. Even a small change in $\gamma$ can drastically change $p^*$. Therefore, in agreement with \cite{1p}, $\gamma$ plays a key role in promoting quantum-cooperation.
\begin{table}[ht]
	\centering
	\caption{Sensitivity indices of $p^*$ with respect to the parameters $D_{g}$, $D_{r}$ and $\gamma$. In this table, we set $D_g=0.9$ and $D_r=0.2$.}\label{T6}
	\begin{tabular}{c|c|c}
		\hline
		$S_{D_g}$ & $S_{D_r}$ & $S_{\gamma}$ \\
		\hline
	  	\begin{tabular}{c|c} $\gamma=\frac{\pi}{9}$ & $\gamma=\frac{\pi}{6}$ \\ \hline $1.029$ & $-0.593$ \\ \end{tabular} & \begin{tabular}{c|c} $\gamma=\frac{\pi}{6}$ & $\gamma=\frac{\pi}{5}$ \\ \hline $-0.173$ & $0.037$ \\ \end{tabular} & \begin{tabular}{c} $\gamma=\frac{\pi}{6}$ \\ \hline $5.596$  \\ \end{tabular} \\
		\hline
	\end{tabular}
\end{table}

In the RDE, both players receive equal expected payoffs, $ \emph{\$}^{*}_{\mathrm{A}(\mathrm{B})} $, such that
\begin{equation}\label{24}
\emph{\$}^{*}_{\mathrm{A}(\mathrm{B})}=(D_{r}-D_{g})p^{*2}+(1-D_{r}+D_{g})p^{*}.
\end{equation}
It is an increasing function of $p^{*}$. From the previous sensitivity analysis, $\gamma$ is the most critical factor in improving the expected payoffs of the RDE. The group benefit, $\emph{\$}^{q}_{\mathrm{A}}+\emph{\$}^{q}_{\mathrm{B}}$ from the RDE is greater than that from the other NEs if
\begin{equation}\label{25}
\sin 2\gamma>\frac{\sqrt{2(D_{r}+2D_{r}D_{g}+D_{g})}}{1+D_{r}+D_{g}}.
\end{equation}

For $\gamma= \arcsin(\sqrt{\frac{D_{r}+D_{g}}{2(1+D_{r}+D_{g})}})$, $p^{*}=0.5$. This means simultaneous quantum-cooperation and defection with the same probability. In this case, the expected payoff of each player is
\begin{equation}\label{26}
\emph{\$}^{q}_{\mathrm{A}(\mathrm{B})}(\hat{U}(0.5)\otimes\hat{U}(0.5))=\frac{2+D_{g}-D_{r}}{4},
\end{equation}
which is the average of the payoff matrix (Table \ref{T4}). This agrees with many limit cases of classical \cite{Satogames,rde1} and quantum \cite{corr1,corr2,decoh1,1p,unique} games.

When $ D_{g}>D_{r} $, within the interval $[\gamma_1,\gamma_2]$, the RDE can provide a fair partial solution. Each player receives  an appropriate payoff. This payoff increases with $\gamma$ and approaches the PO payoff as $\gamma$ approaches $\gamma_2$.

\subsection{The coexistence phase}
If $ D_{g}<D_{r} $, both the classical NE: $\hat{D}\otimes\hat{D}$ and the quantum NE: $\hat{Q}\otimes\hat{Q}$ coexist in the interval $[\gamma_2,\gamma_1]$. The payoffs achieved with $\hat{D}\otimes\hat{D}$ are the worst for both players. Conversely, $\hat{Q}\otimes\hat{Q}$ guarantees the highest payoff for each player. However, each player is concerned that the opponent might deviate to $\hat{D}$. In this situation, a rational player would prefer a safer equilibrium. According to Situ \cite{situ}, the risks associated with both $\hat{D}\otimes\hat{D}$ and $\hat{Q}\otimes\hat{Q}$ are given as follows
\begin{equation}\label{31}
\delta_{\mathrm{A(B)}}(\hat{D}\otimes \hat{D})=0
\end{equation}
\begin{equation}\label{32}
\delta_{\mathrm{A(B)}}(\hat{Q}\otimes \hat{Q})=1+D_{r}-(1+D_{r}+D_{g})\sin^2\gamma.
\end{equation}
Since $\delta_{\mathrm{A(B)}}(\hat{D}\otimes \hat{D})<\delta_{\mathrm{A(B)}}(\hat{Q}\otimes \hat{Q})$, this analysis would tempt players to defect and thus obtain the worst payoff for each player.

Applying the risk-dominance criterion \cite{HS88,rde1}, the total losses due to the deviation from $\hat{Q}\otimes \hat{Q}$ and $\hat{D}\otimes \hat{D}$ are given respectively by Eqs. (\ref{33}) and (\ref{34}).
\begin{equation}\label{33}
\Delta (\hat{Q}\otimes \hat{Q})=[-D_{g}+(1+D_{r}+D_{g})\sin^2\gamma]^2.
\end{equation}
\begin{equation}\label{34}
\Delta (\hat{D}\otimes \hat{D})=[D_{r}-(1+D_{r}+D_{g})\sin^2\gamma]^2.
\end{equation}
Based on the risk-dominance criterion (Eq. (\ref{5})),
\begin{equation}\label{35}
\mathrm{RDE}=\left\{\begin{array}{ll}
\hat{D}\otimes \hat{D}, & \mathrm{if } \gamma_2\leq \gamma<\gamma^*; \\
\hat{U}(0.5)\otimes\hat{U}(0.5), & \mathrm{if } \gamma=\gamma^*; \\
\hat{Q}\otimes \hat{Q}, & \mathrm{if } \gamma^* < \gamma \leq \gamma_1.
\end{array}
\right.
\end{equation}

Now, we examine each case separately. First, if $\gamma_2\leq \gamma<\gamma^*$, the RDE is $\hat{D}\otimes \hat{D}$, and the corresponding payoff is $(0,0)$. If player B deviates from defection, then the expected payoff functions become
\begin{equation}\label{37}
\emph{\$}^{q}_{\mathrm{A}}(\hat{D}\otimes\hat{U}(q))= [1+D_{g}-(1+D_{r}+D_{g})\sin^2\gamma] q,
\end{equation}
and
\begin{equation}\label{38}
\emph{\$}^{q}_{\mathrm{B}}(\hat{D}\otimes\hat{U}(q))= [-D_{r}+(1+D_{r}+D_{g})\sin^2\gamma] q.
\end{equation}
In this interval, $\emph{\$}^{q}_{\mathrm{A}}(\hat{D}\otimes\hat{U}(q))$ is an increasing function of $q$, while $\emph{\$}^{q}_{\mathrm{B}}(\hat{D}\otimes\hat{U}(q))$ is decreasing. The increase in the probability of quantum-cooperation only benefits the defector. This explains why $\hat{D}\otimes \hat{D}$ is the RDE in this interval.

Second, at $\gamma=\gamma^*$, the RDE is $\hat{U}(0.5)\otimes\hat{U}(0.5)$, and the corresponding expected payoff is expressed in Eq. (\ref{26}). If player B deviates from $\hat{U}(0.5)$, then the expected payoff functions become
\begin{equation}\label{39}
\emph{\$}^{q}_{\mathrm{A}}(\hat{U}(0.5)\otimes\hat{U}(q))= q + \frac{D_{g}-D_{r}}{4},
\end{equation}
and
\begin{equation}\label{40}
\emph{\$}^{q}_{\mathrm{B}}(\hat{U}(0.5)\otimes\hat{U}(q))= \frac{2+D_{g}-D_{r}}{4}.
\end{equation}
If $q<0.5$ ($q>0.5$), player A has an unreasonable loss (gain) without affecting player B's expected payoff. Therefore, the fair RDE in this situation is $\hat{U}(0.5)\otimes\hat{U}(0.5)$.

Third, if $\gamma^*\leq \gamma<\gamma_1$, the RDE is $\hat{Q}\otimes \hat{Q}$, and the corresponding payoff is the PO payoff, $(1,1)$. If player B deviates from quantum-cooperation, then the expected payoff functions become
\begin{equation}\label{41}
\emph{\$}^{q}_{\mathrm{A}}(\hat{Q}\otimes\hat{U}(q))= [1+D_{r}-(1+D_{r}+D_{g})\sin^2\gamma] q -D_{r}+(1+D_{r}+D_{g})\sin^2\gamma,
\end{equation}
and
\begin{equation}\label{42}
\emph{\$}^{q}_{\mathrm{B}}(\hat{Q}\otimes\hat{U}(q))= [-D_{g}+(1+D_{r}+D_{g})\sin^2\gamma] q + 1+D_{g}-(1+D_{r}+D_{g})\sin^2\gamma.
\end{equation}
In this interval both $\emph{\$}^{q}_{\mathrm{A}}(\hat{Q}\otimes\hat{U}(q))$ and $\emph{\$}^{q}_{\mathrm{B}}(\hat{Q}\otimes\hat{U}(q))$ are increasing functions of $q$. Then both players benefit from an increase in the level of quantum-cooperation and achieve the PO payoff when player B becomes fully quantum-cooperative.

In the QPD with $D_g<D_r$, if $\gamma=\gamma^*$, then the RDE is $\hat{U}(0.5)\otimes\hat{U}(0.5)$. Each player receives an expected payoff of $\frac{2+D_g-D_r}{4}$. If $\gamma^*<\gamma\leq\gamma_1$, the RDE is $\hat{Q}\otimes\hat{Q}$. Each player receives the PO outcome. Therefore, the QPD is partially solved when $\gamma=\gamma^*$ and fully solved when $\gamma^*<\gamma\leq\frac{\pi}{2}$. In previous studies \cite{Du03,1p}, the QPD is only solved in the interval $[\gamma_1,\frac{\pi}{2}]$.

\section{Conclusions} \label{sec:6}
Little is known about the selection of a unique NE in quantum games, especially in the context of the EWL quantization scheme. We have studied two cases of QPD with different dilemma strength parameters. The first case is characterized by two asymmetric NEs (one player quantum cooperates and the other one defects). In the second case, there are two symmetric NEs: risky PO (mutual quantum-cooperation) and safe non-PO (mutual defection). Both Situ \cite{situ} and the RDE \cite{HS88,rde1} approaches have been applied. The Situ approach failed to find a single safe NE in both cases.

When the gain due to defection if the opponent cooperates is greater than the loss due to cooperation if the opponent
defects and $\gamma_1\leq\gamma\leq\gamma_2$, the RDE is a quantum-mixed-strategy NE. The degree of quantum-cooperation strongly depends on the entanglement and slightly on the dilemma strength parameters. In the RDE, players receive equal payoffs that continuously increase from the punishment to the reward with the degree of quantum-cooperation. For a particular degree of entanglement, the expected payoff of each player due to the RDE is the average of the payoff matrix, which is consistent with many limits of classical \cite{Satogames,rde1} and quantum \cite{corr1,corr2,decoh1,1p,unique} games. Above a certain value of entanglement, the group benefit of the RDE is higher than that of the other equilibria, which is consistent with \cite{rde1}. In this case, the RDE is considered as a partial solution of the PD in its quantum extension.

In the opposite case and when and $\gamma_2\leq\gamma\leq\gamma_1$, the entanglement completely controls the RDE, regardless of the specific values of the dilemma strength parameters. Below a certain threshold, the RDE is the mutual defection, and each player receives the punishment payoff. At the threshold, the RDE is the simultaneous quantum cooperation and defection with the same probability. Each player receives the average of the payoff matrix. Above the threshold, the RDE is the mutual quantum-cooperation, and each player receives the PO outcome. Therefore, the RDE resoles the PD in this interval.

In general, the entanglement completely controls the RDE in QPD. The increase in entanglement increases the probability of quantum-cooperation of the RDE and thus improves its outcome. The selection of the RDE can extend the range in which quantization solves the PD.




\end{document}